\documentclass[conference]{IEEEtran}
\IEEEoverridecommandlockouts

\usepackage{cite}
\usepackage{amsmath,amssymb,amsfonts}
\usepackage{algorithmic}
\usepackage{graphicx}
\usepackage{textcomp}
\usepackage{xcolor}
\usepackage{multirow}

\usepackage{bm}
\usepackage{listings}

\def\BibTeX{{\rm B\kern-.05em{\sc i\kern-.025em b}\kern-.08em
    T\kern-.1667em\lower.7ex\hbox{E}\kern-.125emX}}
\begin{document}

\title{A New Non-Binary Response Generation Scheme \\ from Physical Unclonable Functions
}\vspace{-1em}

\author{\IEEEauthorblockN{Yonghong Bai}
\IEEEauthorblockA{\textit{Broadcom Inc.} \\
Palo Alto, CA, USA \\
yob216@outlook.com}
\and
\IEEEauthorblockN{Zhiyuan Yan}
\IEEEauthorblockA{
\textit{Department of ECE, Lehigh University} \\
Bethlehem, PA, USA \\
zhy6@lehigh.edu}
}

\maketitle
\vspace{-15em}

\begin{abstract}

Physical Unclonable Functions (PUFs) are widely used in key generation, with each PUF cell typically producing one bit of data. To enable the extraction of longer keys, a new non-binary response generation scheme based on the one-probability of PUF bits is proposed. Instead of using PUF bits directly as keys, non-binary responses are first derived by comparing the one-frequency of PUF bits with thresholds that evenly divide the area under the probability density function of the one-probability distribution and then converted to binary keys. To simplify the calculation of these thresholds, a re-scaling process is proposed and the beta distribution is used to model the one-probability distribution. Our FPGA implementation results demonstrate a significant increase in effective key length as opposed to previous works. Finally, we estimate the error rates and biases of the generated keys, and confirm the feasibility of the proposed key generation scheme.
\end{abstract}

\begin{IEEEkeywords}
PUFs, one-probability, non-binary response, FPGA.
\end{IEEEkeywords}
\vspace{-1em}

\section{Introduction}
Machine learning has increasingly found applications across a diverse range of fields, including civil engineering \cite{dan2024evaluation}\cite{dan2024multiple}, security \cite{weng2024big}\cite{weng2024fortifying}, computer vision \cite{zhang2023fully} \cite{gao2024multispectral} \cite{feng2024non}, and graph-based data analysis \cite{liu2024graphsnapshot} \cite{liu2024llmeasyquant}, transforming how these domains approach complex challenges. Integrating physical unclonable functions (PUFs) into machine learning frameworks can significantly enhance the security and efficiency of various machine learning applications. PUFs can be used to generate unique cryptographic keys for encrypting sensitive data utilized in machine learning models. PUF-based keys can authenticate machine learning models to verify their origin and integrity before deployment.

PUFs, embedded in integrated circuits, generate specific outputs by leveraging the inherent mismatches in circuits caused during the manufacturing process, which are uncontrollable and make the PUF outputs unclonable. PUFs are commonly utilized in key generation schemes, typically producing one bit of data per evaluation.However, if PUF bits cannot be reused for multiple key generations, the number of PUF cells required increases proportionally with the number of keys generated,leading to greater hardware overhead, particularly for resourceconstrained
Internet-of-Things devices \cite{zhang2023s}\cite{bavcic2024jy61}.

Recent advancements in machine learning have shown great promise in enhancing the reliability and security of PUFs, which are essential for secure key generation and device authentication. PUFs leverage inherent variations in microfabrication processes to generate unique and unpredictable responses, making them widely applicable for cryptographic applications. However, these responses can be affected by environmental changes and aging, leading to potential reliability issues. Machine learning techniques, particularly those focused on error correction and response stabilization, offer novel approaches to improve the consistency of PUF outputs by adapting to environmental variations and predicting stable response patterns. Additionally, machine learning has been utilized to optimize the design and implementation of PUFs, from selecting the most stable bit patterns to identifying optimal parameters that maximize entropy and minimize bit error rates. As machine learning continues to evolve, its integration with PUF technology promises to significantly enhance PUF resilience and broaden its applications across various secure systems.

Recently, various approaches that aim at extracting more bits from a single PUF cell have been proposed to enhance the efficiency of PUFs. In \cite{suzuki2018efficient}\cite{suzuki2018quaternary}, it is observed that a constant PUF consistently generates either a $0$ or $1$, while an inconstant PUF cell produces values that flip randomly. Ternary and quaternary PUF responses are obtained by assigning specific values to these inconstant outputs. However, these ternary and quaternary responses are primarily used for debiasing rather than directly for key generation. In \cite{bai2021novel}, a quaternary PUF response extraction method based on the one-probability distribution of PUFs was proposed, and it can produce robust quaternary keys with arbitrarily small leakage. However, this method is limited to generating only quaternary keys.

This paper proposes a new generic non-binary response generation scheme based on the one-probability distribution of PUFs. A non-binary response is first produced by comparing the one-frequency of PUF outputs with thresholds that evenly divide the area under the probability density function (PDF) of the PUF's one-probability and then converted to binary keys. This approach is a significant departure from many past works, which typically generate binary keys directly from the outputs of PUF cells. While the method in \cite{bai2021novel} and the proposed scheme are both based on one-probability distribution of PUFs, this work generalizes the method in \cite{bai2021novel} and has the following key contributions:
\begin{enumerate}
\item A re-scaling process is introduced to appropriately normalize the distribution, enabling equal division of the area into sections used for extracting non-binary responses. This helps to alleviate the granularity issue caused by very high density at two extremes for one-probability distribution.
\item Gray codes are used to translate non-binary responses to binary keys so that the responses of adjacent sections differ by only one bit, which minimizes the impact on bit error rates when the one-frequency of a PUF cell shifts between neighboring sections.
\item The one-probability distribution is modeled using the beta distribution, which is then employed to calculate the thresholds. This significantly simplified the calculation of thresholds.
\item Both theoretical analysis and experimental validation of the proposed scheme are presented.  SR-latch PUFs and the proposed non-binary response generation scheme are implemented on an FPGA. The experimental results are used for distribution fitting and performance evaluations, including effective key lengths, error rates, and biases. The evaluation demonstrates a significant increase in effective key length compared to state-of-the-art methods.
\end{enumerate}

\section{Reviews}
\label{Reviews}

Typically, a PUF cell returns a single bit upon random evaluation, and PUF bits are not always perfectly reproducible in practice. Many studies \cite{bai2019secure}\cite{li2022new}\cite{chen2018secret} \cite{gunlu2019code} \cite{hiller2017hiding} \cite{bai2016puf} assume that PUF cells are equally likely to produce a $0$ or a $1$, resulting in a fixed bit error rate. While this assumption simplifies the construction of error correction codes (see, for instance, \cite{bai2017physical}), it does not accurately reflect experimental PUF behavior, where some PUF cells are more likely to consistently produce stable $0$'s or $1$'s, while others frequently toggle between $0$ and $1$ \cite{maes2013accurate}.

\begin{figure}[h]
\centering
\includegraphics[width=3.1in]{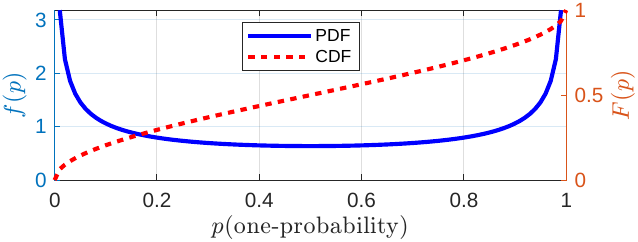}
\caption {The probability density function and cumulative density function of the one-probability distribution.}
\label{betaDist}
\end{figure}

The one-probability of a PUF cell is defined to be the likelihood that a PUF cell outputs a $1$ upon random evaluation. Experimental observations show that each PUF cell has its unique one-probability. In \cite{maes2013accurate}, the one-probabilities are modeled as random samples from a random variable $P$, with its probability density function (PDF) and cumulative distribution function (CDF) illustrated in \mbox{Fig.~\ref{betaDist}}. Experimental results show that most PUF cells are more likely to consistently generate stable 0 or 1 outputs, while others frequently alternate between 0 and 1 across multiple evaluations. This results in a higher one-probability density at the two extremes, creating a U-shaped distribution. as shown in Fig.~\ref{betaDist}.

\vspace{-1em}

\section{Non-binary Response Extraction}
\label{nonBinExtract}

In this section, we describe how our proposed non-binary response generation scheme operates, using the one-probability as the basis for generating non-binary responses rather than the direct output of the PUF. Instead of extracting binary responses directly from a PUF cell output,  we extract non-binary responses from PUFs based on the one-probability distribution of PUF cells, which is shown in Fig. \ref{betaDist}.


The region under the PDF of the one-probability is evenly divided into $2^t$ individual sections using $2^t-1$ thresholds to extract $t$-ary responses. These sections are indexed as $0, 1, \dots, 2^t-1$. The non-binary extraction of a PUF corresponds to the index of the section identified by its one-probability. Ideally, these sections should have equal areas to ensure that a one-probability falls into each section with the same likelihood, minimizing the bias of the extracted non-binary responses. In practice, we apply one-frequency to approximate the one-probability of PUFs for extracting non-binary responses. We collect $k$ outputs from each PUF cell. If the number of $1$'s observed in the $k$ trials is $m$, the one-frequency of the PUF cell is $m/k$. If the one-frequency falls within section $i$, the non-binary response of the PUF cell is $i$, which corresponds to the index of that section. Because the possible one-frequencies are discrete, it is possible that the area cannot be evenly divided into $2^t$ sections.

However, the density at the two extremes ($0$ and $1$) is significantly higher than that in the middle, and this results in
some thresholds very close to 0 and 1, requiring more observations to distinguish between the sections. Otherwise, the bias between the extracted values will be substantial. To mitigate this issue, we re-scale the collected data to reduce the bias in the extracted responses as well as the number of observations needed:
\begin{enumerate}
\item First, we separate all-zero and all-one observations from the rest of the data. If a PUF cell produces all-zero or all-one bits in $k$ observations, it probably generate those bits stably. These stable PUF cells can be used for key generation and authentication with minimal or even no error correction.

\item The remaining PUF cells are used to extract non-binary responses. We begin by identifying the minimum and maximum one-frequencies from the remaining data. Then, we compute $2^t-1$ vertical lines that divide the region under the PDF, spanning from the minimum to the maximum one-frequencies, into $2^t$ separate areas as evenly as possible. These vertical line positions are used as thresholds for extracting $t$-ary responses.
\end{enumerate}

\begin{figure}[h]
\centering
\includegraphics[width=3.1in]{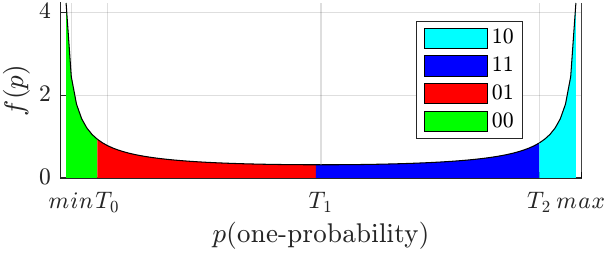}
\caption {Quaternary response extraction.}
\label{quaternary}
\end{figure}

For example, in quaternary response extraction, three thresholds $T_0$, $T_1$, and $T_2$ divide the one-frequency range. The quaternary response will be $0$, $1$, $2$, or $3$ if the one-frequency of the PUF cell falls into the intervals $[min, T_0)$,  $[T_0, T_1)$, $[T_2,T_3)$, and $[T_3,max]$, respectively, as shown in Fig. \ref{quaternary}. Note that $min$ and $max$ would be $0$ and $1$, respectively, without re-scaling.

In the proposed scheme, non-binary (power of $2$) responses are mapped to sequences of bits. An error occurs when the one-frequency of a PUF cell, upon different evaluations, falls into different sections. Assuming the one-probability of a PUF cell remains constant, the one-frequency distribution follows a binomial distribution. This means that the likelihood of the one-frequency landing in neighboring sections is higher than landing in more distant sections. To reduce the bit error rate, Gray codes are used in our scheme to map non-binary responses to bits. Gray codes are designed so that adjacent values differ by only one bit. As a result, when the one-frequency crosses into an adjacent section, only a single bit of the generated sequence of bits will flip, minimizing the overall impact on the bit error rate. For example, in Fig. \ref{quaternary}, the Gray codes for the quaternary symbols $0$, $1$, $2$, and $3$ are `$00$', `$01$', `$11$', and `$10$', respectively.

The one-probability distribution model in \cite[Eq. (2)]{maes2013accurate} is too complex for the threshold calculation and parameter estimation. Instead, we model the one-probability distribution using the beta distribution \cite{mackay2003information}, a well-known continuous probability distribution defined over the interval $[0,1]$. The beta distribution is particularly suitable for modeling probabilities or percentages, such as the one-probability considered in this paper. The PDF of the beta distribution for $p \in [0,1]$ is expressed as:
\begin{equation}
f(p) = \frac{p^{\alpha-1}(1-p)^{\beta-1}}{B(\alpha,\beta)},
\label{beta_pdf}
\end{equation}
where $p$ is an observed value of a random variable $P$, $\alpha$ and $\beta$ are shape parameters, and $B(\alpha,\beta)$ is the beta function, a normalization constant that ensures the total probability integrates to 1.

Larger alphabets, such as 16-ary, result in longer responses but also increase complexity. To ensure accurate extraction, more PUF evaluations are required to correctly position thresholds across the probability distribution. Additionally, using larger alphabets tends to increase symbol and bit error rates, which in turn necessitate advanced error-correction techniques or additional evaluations to achieve reliable key generation. This tradeoff between key length and reliability must be carefully managed to maintain the efficiency and security of the system.

\section{FPGA Implementation and Discussion}
\label{FPGADes}

\begin{table*}[h]
\renewcommand{\arraystretch}{1.1}
\caption{Comparison with state-of-the-arts based on our FPGA implementation of $1024$ SR-latch PUFs. The symbol error rate applies to only the $55$ PUF cells generating non-binary outputs, while the other metrics refer to all $1024$ PUF cells.}
\centering
\begin{tabular}{ |c|c|c|c|c|c|c|}
\hline

{}& {Alphabet Size} & {Entropy} & {Effective Key Length} & {Symbol Error Rate}  & {Bit Error Rate} & {Bias} \\ \hline

 {\multirow{3}{*} {This work}} & {Quaternary} & {$1.0479$} & {$1073$} & {$0.0980$} & {$0.0050$} & {$0.5403$}\\  \cline{2-7}

 {} & {$8$-ary} & {$1.0977$} & {$1124$} & {$0.2413$}  & {$0.0117$} & {$0.5344$}\\ \cline{2-7}

 {} & {$16$-ary} & {$1.1484$} & {$1176$} & {$0.4773$} & {$0.0221$} & {$0.5315$}\\  \hline
 {Binary works \cite{chen2018secret}\cite{maes2015secure}\cite{ueno2019tackling}\cite{bai2021secure}} & {Binary} & {$\leq 0.9962$} & {$\leq 1020$} & {-} & {$0.0043$} & {$0.5361$} \\ \hline

\end{tabular}
\label{thresholdsDistr}
\end{table*}

\subsection{FPGA implementation of SR-latch PUFs}

Many secure designs are implemented on FPGA platforms \cite{9517289}. In our experiments, $1024$ SR-latch PUFs are implemented on a Xilinx Kintex-7 FPGA (XC7K325T-2FFG900C).
Each PUF cell was evaluated $1048575$ times. The experimental data show that out of $1024$ PUF cells, $549$ produced more than half $1$'s, and $475$ produced less than half $1$'s across the total observations. Notably, $520$ PUF cells consistently produced a $1$ in all $1048575$ evaluations, while $449$ PUF cells produced all $0$'s. These stable PUF cells are used for key generation directly with minimal error correction, as their output is highly predictable. For the remaining $55$ PUF cells, which exhibited more variability in their output (neither always $1$'s nor always $0$'s), we applied our re-scaling technique to extract non-binary responses.

The estimated parameters of the beta distribution based on our experimental results are $\hat{\alpha} = 0.0032$ and $\hat{\beta} = 0.0028$. For non-binary response extraction, we first re-scale the one-frequencies, where $min$ and $max$ are set to 1/1048575 and 1048574/1048575, respectively, regardless of the non-binary responses.
For quaternary responses, the possible values are $0$, $1$, $2$, and $3$. The thresholds used to distinguish between these values are \{$0.0010616$, $0.5049029$, $0.998969$\}. We generated quaternary responses by comparing the one-frequencies of each PUF cell against these thresholds and the re-scaled $min$ and $max$ values. The distribution of quaternary responses among the $55$ PUF cells is \{$10$, $16$, $18$, $11$\}. Similarly, the thresholds for 8-ary and 16-ary response extractions are \{$0.000032$, $0.001061$, $0.032387$, $0.504902$, $0.968752$, $0.998969$, $0.999968$\} and \{$0.000005$, $0.000032$, $0.000186$, $0.001061$, $0.005956$, $0.032387$, $0.156357$, $0.504902$, $0.848678$, $0.968752$, $0.994241$, $0.998969$, $0.999817$, $0.999968$, $0.999994$\}, respectively. The distributions of the 8-ary and 16-ary responses among the $55$ PUF cells are \{$6$, $4$, $4$, $12$, $10$, $8$, $7$, $4$\} and \{$3$, $3$, $3$, $1$, $2$, $2$, $4$, $8$, $4$, $6$, $4$, $4$, $4$, $3$, $1$, $3$\}.

\subsection{Comparison with state-of-the-arts}
We compare the performance of the proposed scheme with previous key generation schemes \cite{chen2018secret} \cite{maes2015secure} \cite{ueno2019tackling} \cite{bai2021secure}, in terms of entropy, key length, error rates, and biases. To provide a fair comparison, we assume the $1024$ SR-latch PUFs implemented in FPGA are the PUFs used by all key generation schemes. Another advantage of this comparison is that the performance metrics are based on actual PUF implementations. As mentioned above, the key generation scheme in \cite{bai2021novel} will require much more evaluations than the $1048575$ evaluations mentioned above to work without re-scaling. Hence, the comparison herein does not include the key generation scheme in \cite{bai2021novel}.

Since the primary motivation of this work is to extract longer keys from PUFs, we compare the key lengths achieved by our method with those of state-of-the-art binary key generation techniques in \cite{chen2018secret}\cite{maes2015secure}\cite{ueno2019tackling}\cite{bai2021secure}. We adopt the effective key length, instead of key length, as the metric, since it accounts for the bias in the key (and hence the leakage). Therefore, the effective key length is a more relevant metric for security. In the binary case, if the number of $1$’s in $1048575$ evaluations exceeds half (i.e., $1048575/2$), the binary response is ‘1’; otherwise, it is ‘0’. Based on this approach, $549$ PUF cells generate 1’s, while the remaining $475$ generate 0’s. To compute the effective key length, we approximate the fraction of 1’s (or 0’s) as the probability of generating a ‘1’ (or ‘0’) in a binary string. The entropy $H_1 = -\frac{549}{1024}\log_2{\frac{549}{1024}}-\frac{475}{1024}\log_2{\frac{475}{1024}} = 0.9962$. This implies that up to $H_1 \times 1024 = 1020$ bits can be derived from the $1024$ PUF bits. In this work, stable PUF bits (i.e., bits from cells that consistently produce either 0’s or 1’s) are used to derive keys without error correction, while the remaining cells generate non-binary responses. The entropy of the stable PUF bits is $H_2 = -\frac{449}{969}\log_2{\frac{449}{969}}-\frac{520}{969}\log_2\frac{520}{969} = 0.9961$. For the non-binary responses extracted from the $55$ PUFs, the entropies are $1.9571$ for quaternary, $2.8832$ for 8-ary, and $3.8307$ for 16-ary response. Hence, the entropy of quaternary, 8-ary, and 16-ary response extractions, with respect to all $1024$ PUF cells, is $H_2 \times 969/1024 + 1.9571 \times 55/1024 = 1.0479$, $H_2 \times 969/1024 + 2.8832 \times 55/1024 = 1.0977$, and $H_2 \times 969/1024 + 3.8307 \times 55/1024 = 1.1484$, respectively. The effective key lengths for quaternary, 8-ary, and 16-ary response extractions are $1073$, $1124$, and $1176$, respectively. Among the binary key generation methods in \cite{chen2018secret}\cite{maes2015secure}\cite{ueno2019tackling}\cite{bai2021secure}, the entropy per PUF cell is no more than $0.9962$ and the effective key length is no more than $1020$.   These results show that our non-binary response extraction yields significantly longer keys than these previous works.


A theoretical analysis of the error rate for non-binary responses is complex, and hence we rely on experimental results for evaluation. We conducted $100$ experiments using the FPGA implementation, collecting $1048575$ PUF outputs in each experiment to extract non-binary responses. One of the experiment results was randomly selected as a benchmark (the enrolled response) and used to estimate the beta distribution parameters and evaluate the error rate. A symbol error occurs when the non-binary response from an experiment differs from that of the benchmark. The experimental results indicate the symbol error rates for quaternary, 8-ary, and 16-ary responses are $0.0980$, $0.2413$, and $0.4773$, respectively, among the $55$ PUF cells. Non-binary error correction codes (ECCs) are often used to correct symbol errors; for instance, quaternary polar codes were employed in \cite{bai2021novel} to correct errors in quaternary keys. However, designing distinct non-binary ECCs for different key sizes increases design complexity. In this work, we encode the symbols using binary codes and use binary ECCs to correct bit errors, simplifying the design process. In our experiment, all observed errors correspond to adjacent values of the benchmark response, meaning that all symbols in error are neighboring symbol values. Therefore, after employing Gray codes, the bit error rate is reduced to $1/\log_2t$ of the symbol error rate for $t$-ary responses, with values of $0.0490$, $0.0804$, and $0.1193$ for quaternary, 8-ary, and 16-ary responses, respectively, among the $55$ PUF cells. The observed bit error rate for the 969 stable PUF cells is $2.915 \times 10^{-8}$. Therefore, when considering the stable PUF bits along with the extracted non-binary bits, the bit error rates for quaternary, 8-ary, and 16-ary extractions are $0.0050$, $0.0117$, and $0.0221$, respectively.

As shown in Table \ref{thresholdsDistr}, the symbol and bit error rates increase with larger alphabets. This happens because as the sections dividing the one-probability distribution become denser, the likelihood of the one-frequency landing in the wrong section increases, resulting in higher error rates. The smaller separation between thresholds leads to greater sensitivity, making it more likely for minor fluctuations in the PUF response to cause symbol errors.

If the $1024$ SR-latch PUFs are used for binary key generation following the methods in \cite{chen2018secret}\cite{maes2015secure}\cite{ueno2019tackling}\cite{bai2021secure}, the average bit error rate will be $0.0043$ per PUF bit. As shown in \mbox{Table \ref{thresholdsDistr}}, the bit error rate in this work is higher compared to that of binary key generation methods.

The bias in the binary strings generated from non-binary responses represents the fraction of 1's. Since these binary strings are used for key generation, an ideal bias of $0.5$ is desired to prevent information leakage, as explained in \cite{ignatenko2010information}. The experimental results in our work show biases of $0.5727$, $0.5212$, and $0.5091$ for quaternary, 8-ary, and 16-ary response extractions, respectively. As the alphabet size increases, the bias decreases. Using the $1024$ SR-latch PUFs, the binary key generation methods in \cite{chen2018secret}\cite{maes2015secure}\cite{ueno2019tackling}\cite{bai2021secure} will result in a bias of $0.5361$.

\subsection{Discussion about the feasibility of the proposed scheme}
The previous works \cite{chen2018secret}\cite{maes2015secure}\cite{ueno2019tackling}\cite{bai2021secure} have explored error correction techniques for PUFs with varying bit error rates. The bit error rates reported in those studies range from $0$ to $0.3$. For example, in \cite{chen2018secret}, polar codes were used to achieve a key generation failure rate of $\le 10^{-6}$ when the bit error rate was $0.3$. This suggests that with appropriate error correction methods, the bit error rates observed in our experiments can be managed effectively for secure key generation.

To mitigate the information leakage caused by these non-ideal biases, various methods can be applied, including Von Neumann's debiasing technique \cite{maes2015secure}, ternary approach \cite{suzuki2018efficient}, and wiretap polar codes \cite{bai2021secure}. In \cite{bai2021secure}, the secrecy leakage from biases as high as $0.9$ could be reduced to a negligible amount by employing OR-based debiasing and wiretap polar codes. Thus, similar approaches can be effectively applied here to minimize the leakage and ensure secure key generation despite the observed bias levels.

\section{Conclusion}
In this paper, we propose a new non-binary response generation scheme for PUFs, where the one-frequency of PUF outputs is compared against thresholds that evenly divide the area under the PDF of the one-probability distribution.  
The one-probability distribution is skewed toward the two extremes, leading to a significant number of all-zero or all-one outputs and making threshold calculations difficult. To address this, a re-scaling process is proposed to use stable PUF cells for key generation directly and adjust the remaining data for non-binary response generation. Gray codes are applied to convert non-binary responses to bits, reducing bit error rates by ensuring that adjacent sections differ by only one bit, thereby minimizing the effect of small shifts in the one-frequency of PUF cells.

The one-probability distribution is represented by the beta distribution, which is subsequently used to determine the thresholds for extracting non-binary responses after estimating its parameters.

We validated our design by implementing SR-latch PUFs and our non-binary response generation scheme on a Xilinx FPGA. 
Our experimental results demonstrated a significant increase in effective key length when compared with previous works. Analysis of the bit error rate and bias in the non-binary responses confirmed the feasibility of our scheme for key generation applications.

\bibliographystyle{IEEEtran}
\bibliography{reference}

\begin{thebibliography}{10}
\providecommand{\url}[1]{#1}
\csname url@samestyle\endcsname
\providecommand{\newblock}{\relax}
\providecommand{\bibinfo}[2]{#2}
\providecommand{\BIBentrySTDinterwordspacing}{\spaceskip=0pt\relax}
\providecommand{\BIBentryALTinterwordstretchfactor}{4}
\providecommand{\BIBentryALTinterwordspacing}{\spaceskip=\fontdimen2\font plus
\BIBentryALTinterwordstretchfactor\fontdimen3\font minus
  \fontdimen4\font\relax}
\providecommand{\BIBforeignlanguage}[2]{{%
\expandafter\ifx\csname l@#1\endcsname\relax
\typeout{** WARNING: IEEEtran.bst: No hyphenation pattern has been}%
\typeout{** loaded for the language `#1'. Using the pattern for}%
\typeout{** the default language instead.}%
\else
\language=\csname l@#1\endcsname
\fi
#2}}
\providecommand{\BIBdecl}{\relax}
\BIBdecl

\bibitem{dan2024evaluation}
H.-C. Dan, B.~Lu, and M.~Li, ``Evaluation of asphalt pavement texture using
  multiview stereo reconstruction based on deep learning,'' \emph{Construction
  and Building Materials}, vol. 412, p. 134837, Jan. 2024.

\bibitem{dan2024multiple}
H.-C. Dan, P.~Yan, J.~Tan, Y.~Zhou, and B.~Lu, ``Multiple distresses detection
  for asphalt pavement using improved you only look once algorithm based on
  convolutional neural network,'' \emph{International Journal of Pavement
  Engineering}, vol.~25, no.~1, p. 2308169, Dec. 2024.

\bibitem{weng2024big}
Y.~Weng, ``Big data and machine learning in defence,'' \emph{International
  Journal of Computer Science and Information Technology}, vol.~16, no.~2, pp.
  25--35, Mar. 2024.

\bibitem{weng2024fortifying}
Y.~Weng, J.~Wu \emph{et~al.}, ``Fortifying the global data fortress: a
  multidimensional examination of cyber security indexes and data protection
  measures across 193 nations,'' \emph{International Journal of Frontiers in
  Engineering Technology}, vol.~6, no.~2, pp. 13--28, Mar. 2024.

\bibitem{zhang2023fully}
H.~Zhang, K.~Gao, H.~Huang, S.~Hou, J.~Li, and G.~Wu, ``Fully decouple
  convolutional network for damage detection of rebars in \mbox{RC} beams,''
  \emph{Engineering Structures}, vol. 285, p. 116023, Jun. 2023.

\bibitem{gao2024multispectral}
K.~Gao, H.~Zhang, and G.~Wu, ``A multispectral vision-based machine learning
  framework for non-contact vehicle weigh-in-motion,'' \emph{Measurement}, vol.
  226, p. 114162, Mar. 2024.

\bibitem{feng2024non}
J.~Feng, K.~Gao, H.~Zhang, W.~Zhao, G.~Wu, and Z.~Zhu, ``Non-contact vehicle
  weight identification method based on explainable machine learning models and
  computer vision,'' \emph{Journal of Civil Structural Health Monitoring}, pp.
  1--18, Mar. 2024.

\bibitem{liu2024graphsnapshot}
D.~Liu, R.~Waleffe, M.~Jiang, and S.~Venkataraman, ``Graphsnapshot: Graph
  machine learning acceleration with fast storage and retrieval,'' \emph{arXiv
  preprint arXiv:2406.17918}, Jun. 2024.

\bibitem{liu2024llmeasyquant}
D.~Liu, M.~Jiang, and K.~Pister, ``\mbox{LLMEasyQuant--An} easy to use toolkit
  for \mbox{LLM} quantization,'' \emph{arXiv preprint arXiv:2406.19657}, Jun.
  2024.

\bibitem{zhang2023s}
Y.~Zhang, C.~Slocum, J.~Chen, and N.~Abu-Ghazaleh, ``It's all in your head
  (set): Side-channel attacks on $\{$AR/VR$\}$ systems,'' in \emph{32nd USENIX
  Security Symposium (USENIX Security 23)}, 2023, pp. 3979--3996.

\bibitem{bavcic2024jy61}
B.~Ba{\v{c}}i{\'c}, C.~Feng, and W.~Li, ``Jy61 imu sensor external validity: A
  framework for advanced pedometer algorithm personalisation,'' \emph{ISBS
  Proceedings Archive}, vol.~42, no.~1, p.~60, July 2024.

\bibitem{suzuki2018efficient}
M.~Suzuki, R.~Ueno, N.~Homma, and T.~Aoki, ``Efficient fuzzy extractors based
  on ternary debiasing method for biased physically unclonable functions,''
  \emph{IEEE Transactions on Circuits and Systems I: Regular Papers}, vol.~66,
  no.~2, pp. 616--629, Sep. 2018.

\bibitem{suzuki2018quaternary}
------, ``Quaternary debiasing for physically unclonable functions,'' in
  \emph{2018 IEEE 48th International Symposium on Multiple-Valued Logic}.\hskip
  1em plus 0.5em minus 0.4em\relax IEEE, May 2018, pp. 7--12.

\bibitem{bai2021novel}
Y.~Bai and Z.~Yan, ``A novel key generation scheme using quaternary puf
  responses and wiretap polar coding,'' \emph{IEEE Communications Letters},
  vol.~25, no.~7, pp. 2142--2145, Apr. 2021.

\bibitem{bai2019secure}
------, ``A secure and robust key generation method using physical unclonable
  functions and polar codes,'' in \emph{2019 IEEE International Workshop on
  Signal Processing Systems (SiPS)}.\hskip 1em plus 0.5em minus 0.4em\relax
  IEEE, Oct. 2019, pp. 254--259.

\bibitem{li2022new}
G.~Li, K.~T. Mursi, A.~O. Aseeri, M.~S. Alkatheiri, and Y.~Zhuang, ``A new
  security boundary of component differentially challenged \mbox{XOR PUFs}
  against machine learning modeling attacks,'' \emph{arXiv preprint
  arXiv:2206.01314}, Jun. 2022.

\bibitem{chen2018secret}
B.~Chen and F.~M. Willems, ``Secret key generation over biased physical
  unclonable functions with polar codes,'' \emph{IEEE Internet of Things
  Journal}, vol.~6, no.~1, pp. 435--445, Aug. 2018.

\bibitem{gunlu2019code}
O.~G{\"u}nl{\"u}, O.~{\.I}{\c{s}}can, V.~Sidorenko, and G.~Kramer, ``Code
  constructions for physical unclonable functions and biometric secrecy
  systems,'' \emph{IEEE Transactions on Information Forensics and Security},
  vol.~14, no.~11, pp. 2848--2858, Apr. 2019.

\bibitem{hiller2017hiding}
M.~Hiller and A.~G. {\"O}nalan, ``Hiding secrecy leakage in leaky helper
  data,'' in \emph{International Conference on Cryptographic Hardware and
  Embedded Systems}.\hskip 1em plus 0.5em minus 0.4em\relax Springer, Sep.
  2017, pp. 601--619.

\bibitem{bai2016puf}
Y.~Bai, L.~Wu, X.~Wu, X.~Li, X.~Zhang, and B.~Wang, ``{PUF}-based encryption
  method for {IC} cards on-chip memories,'' \emph{Electronics Letters},
  vol.~52, no.~20, pp. 1671--1673, Sep. 2016.

\bibitem{bai2017physical}
Y.~Bai and Z.~Yan, ``Physical unclonable functions with improved robustness
  based on polar codes,'' in \emph{Signal Processing Systems (SiPS), 2017 IEEE
  International Workshop on}.\hskip 1em plus 0.5em minus 0.4em\relax IEEE, Oct.
  2017, pp. 1--6.

\bibitem{maes2013accurate}
R.~Maes, ``An accurate probabilistic reliability model for silicon
  \mbox{PUFs},'' in \emph{International Workshop on Cryptographic Hardware and
  Embedded Systems}.\hskip 1em plus 0.5em minus 0.4em\relax Springer, Aug.
  2013, pp. 73--89.

\bibitem{mackay2003information}
D.~J. MacKay, D.~J. Mac~Kay \emph{et~al.}, \emph{Information theory, inference
  and learning algorithms}.\hskip 1em plus 0.5em minus 0.4em\relax Cambridge
  university press, 2003.

\bibitem{maes2015secure}
R.~Maes, V.~van~der Leest, E.~van~der Sluis, and F.~Willems, ``Secure key
  generation from biased \mbox{PUFs},'' in \emph{International Workshop on
  Cryptographic Hardware and Embedded Systems}.\hskip 1em plus 0.5em minus
  0.4em\relax Springer, Sep. 2015, pp. 517--534.

\bibitem{ueno2019tackling}
R.~Ueno, M.~Suzuki, and N.~Homma, ``Tackling biased \mbox{PUFs} through biased
  masking: A debiasing method for efficient fuzzy extractor,'' \emph{IEEE
  Transactions on Computers}, vol.~68, no.~7, pp. 1091--1104, July 2019.

\bibitem{bai2021secure}
Y.~Bai and Z.~Yan, ``A secure and robust \mbox{PUF-based} key generation with
  wiretap polar coset codes,'' \emph{Journal of Electronic Testing}, vol.~37,
  no.~3, pp. 305--316, Jun. 2021.

\bibitem{9517289}
Y.~Zhang, R.~Yasaei, H.~Chen, Z.~Li, and M.~A.~A. Faruque, ``Stealing neural
  network structure through remote \mbox{FPGA} side-channel analysis,''
  \emph{IEEE Transactions on Information Forensics and Security}, vol.~16, pp.
  4377--4388, Aug. 2021.

\bibitem{ignatenko2010information}
T.~Ignatenko and F.~M. Willems, ``Information leakage in fuzzy commitment
  schemes,'' \emph{IEEE Transactions on Information Forensics and Security},
  vol.~5, no.~2, pp. 337--348, Jun. 2010.

\end{thebibliography}

\end{document}